\documentclass[floatfix,reprint,prl,aps,twocolumn,superscriptaddress,longbibliography]{revtex4-2}
\usepackage[bitstream-charter]{mathdesign}
\usepackage{float}
\usepackage{amsmath}
\DeclareMathOperator{\sign}{sgn}
\usepackage{graphicx}
\usepackage{dcolumn}
\usepackage{bm}
\usepackage{calrsfs}
\newcommand{\K}{{\bf k}} 

\begin{document}

\title{Universal composite boson formation in strongly interacting one-dimensional fermionic systems}

\author{Francesc Sabater}
\affiliation{Departament de F{\'i}sica Qu{\`a}ntica i Astrof{\'i}sica, Facultat de F{\'i}sica, Universitat de Barcelona, E-08028 Barcelona, Spain}
\affiliation{Institut de Ci{\`e}ncies del Cosmos, Universitat de Barcelona, ICCUB, Mart{\'i} i Franqu{\`e}s 1, E-08028 Barcelona, Spain}
\author{Abel Rojo-Francàs}
\affiliation{Departament de F{\'i}sica Qu{\`a}ntica i Astrof{\'i}sica, Facultat de F{\'i}sica, Universitat de Barcelona, E-08028 Barcelona, Spain}
\affiliation{Institut de Ci{\`e}ncies del Cosmos, Universitat de Barcelona, ICCUB, Mart{\'i} i Franqu{\`e}s 1, E-08028 Barcelona, Spain}
\author{Grigori E. Astrakharchik}
\affiliation{Departament de F\'isica, Universitat Polit\`ecnica de Catalunya, 
Campus Nord B4-B5, E-08034 Barcelona, Spain}
\affiliation{Departament de F{\'i}sica Qu{\`a}ntica i Astrof{\'i}sica, Facultat de F{\'i}sica, Universitat de Barcelona, E-08028 Barcelona, Spain}
\affiliation{Institut de Ci{\`e}ncies del Cosmos, Universitat de Barcelona, ICCUB, Mart{\'i} i Franqu{\`e}s 1, E-08028 Barcelona, Spain}
\author{Bruno Juli{\'a}-D{\'i}az}
\affiliation{Departament de F{\'i}sica Qu{\`a}ntica i Astrof{\'i}sica, Facultat de F{\'i}sica, Universitat de Barcelona, E-08028 Barcelona, Spain}
\affiliation{Institut de Ci{\`e}ncies del Cosmos, Universitat de Barcelona, ICCUB, Mart{\'i} i Franqu{\`e}s 1, E-08028 Barcelona, Spain}

\date{\today}

\begin{abstract}
Attractive $p$-wave one-dimensional fermions are studied in the fermionic Tonks-Girardeau regime in which the diagonal properties are shared with those of an ideal Bose gas. 
We study the off-diagonal properties and present analytical expressions for the eigenvalues of the one-body density matrix. 
One striking aspect is the universality of the occupation numbers which are independent of the specific shape of the external potential. 
We show that the occupation of natural orbitals occurs in pairs, indicating the formation of composite bosons, each consisting of two attractive fermions. 
The formation of composite bosons sheds light on the pairing mechanism of the system orbitals, yielding a total density equal to that of a Bose-Einstein condensate. 
\end{abstract}
                             
\maketitle

\textbf{Introduction.} 
One of the most fascinating quantum phenomena is Bose-Einstein condensation (BEC), characterized by a macroscopic occupation of a single quantum state~\cite{PitaevskiiStringariBook}. If BEC occurs, the largest eigenvalue of the one-body density matrix (OBDM) $\rho_1({\bf r},{\bf r'})=\langle\hat \Psi^\dagger({\bf r})\hat\Psi({\bf r'})\rangle$ is proportional to the total number of particles $N$ where $\hat\Psi^\dagger({\bf r})$ ($\hat\Psi({\bf r})$) are creation (annihilation) operators of a particle at position ${\bf r}$, respectively~\cite{Lowdin55,PenroseOnsager56}. 
For a homogeneous gas, the corresponding eigenvector of the OBDM is the zero-momentum state which gives rise to the presence of Off-Diagonal Long-Range Order (ODLRO), $\lim_{|{\bf r}-{\bf r}'|\to\infty}\rho_1({\bf r},{\bf r}')\to n_0$ where $n_0$ is the condensate density. 
In stark contrast, due to the Pauli principle, no two fermions are allowed to have the same quantum numbers, and atomic condensation is prohibited. 
Instead, in this Letter we discuss and explore the possibility that strongly attractive fermions form composite bosons, whose density is equal to that of a Bose-Einstein condensate. 

One-dimensional geometry is very special as due to Girardeau's mapping, bosonic and fermionic systems might possess exactly the same diagonal properties~\cite{girardeau_mapping,Cheon1999}. 
In that way, the wave function of impenetrable bosons can be mapped to the wave function of ideal fermions, known as Tonks-Girardeau (TG) gas~\cite{girardeau_mapping,Girardeau2001}.
As well, single-component fermions with a strong $p$-wave attraction can exhibit identical diagonal correlations as in an ideal Bose gas, resulting in what is known as the fermionic Tonks-Girardeau (FTG) gas~\cite{GirardeauNguyenOlshanii2004,GirardeauMinguzzi2006,MinguzziGirardeau2006,GirardeauAstrakharchik2010}. 
Recent experimental advances in spin-polarized fermions with $p$-wave resonances pave the way for the experimental realization of the FTG gas using confinement-induced resonances~\cite{Granger2004,Chang2020}. 
Another peculiarity of the one-dimensional geometry is that quantum fluctuations destroy BEC in gases with finite interactions. 
Indeed, in uniform systems, the ODLRO vanishes in power-law decay $\lim_{|x-x'|\to\infty}\rho_1(x,x')\to n / (n|x-x'|)^{1/(2K)}$ where $K$ is the Luttinger parameter~\cite{Haldane81,Cazalilla_2004}. 
Formally, the ODLRO is restored in the limit of ideal Bose gas ($K\to \infty$). 
Its fermionic counterpart corresponds to the FTG gas which exhibits partial pair condensation as has been discussed in the case of harmonic trapping~\cite{GirardeauMinguzzi2006}.
Recently, Ko\'sik and Sowi\'nski have proven that in FTG the eigenvalues of the OBDM are independent of the shape of the trapping potential~\cite{koscik}. 
This fact enables us to conduct a completely general and universal study of the FTG gas without the need to constrain the study to a specific system or confining potential, as has been done in previous works \cite{MinguzziGirardeau2006}.

In this Letter we analytically study the coherence in a finite-size FTG state in an arbitrary external field. 
We demonstrate that in FTG gas, fermions pair into composite bosons. 
As a consequence, each fermionic pair exhibits a density profile equivalent to that of an ideal boson.
To demonstrate that, we perform natural orbital analysis diagonalizing the one-body density matrix and deduce analytical expressions for its occupation numbers and natural orbitals.
We verify the correctness of the obtained expressions by comparison with numerical diagonalization. 
For a large number of particles, we find that the occupation of the OBDM asymptotically approaches a Lorentzian shape, the same as the one found in Ref.~\cite{BenderErkerGranger2005} for the momentum distribution of a homogeneous FTG gas in the thermodynamic limit. 
It is remarkable that eigenvalues are universal and remain exactly the same for any shape of the external potential~\cite{koscik}.
In particular, for an untrapped system, the eigenvalues of the OBDM correspond to the momentum distribution allowing us to obtain its exact expression. 
Moreover, we demonstrate that the eigenvalues of the OBDM come in pairs. 
In that way, while all the individual fermionic eigenstates of the OBDM are different, the density of fermionic pairs (i.e. composite bosons) remains exactly the same, although its specific shape depends on the external potential. 
Thus, we provide a comprehensible picture of the mechanism of composite boson formation and how this results in the density profile being equal to that of a Bose-Einstein condensate.

\textbf{Fermionic Tonks-Girardeau gas.}
FTG gas describes $N$ fermions with coordinates $x_1,\cdots,x_N$ interacting via short-range $p$-wave attraction tuned in such a way that the ground state wave function 
\begin{equation}
\label{gs atractive fermions} 
\psi_F(x_1,\dots ,x_N)=\psi_B(x_1,\dots ,x_N)\prod\limits_{j<k}^N \sign{(x_k-x_j)}\hspace{2pt},
\end{equation}
is related to the one of an ideal Bose gas, $\psi_B(x_1,\dots,x_N)=\prod_{i=1}^{N}\phi_0(x_i)$, according to Girardeau's mapping~\cite{girardeau_mapping} where $\phi_0(x)$ is the single-particle ground state. 
The coherence properties are encoded in the OBDM,
\begin{equation*}
\rho_1(\!x,x\prime)=N\!\!\int\!\!\!\psi_F(\!x,x_2,\hdots x_N)\psi_F^*(\!x\prime,x_2, \hdots x_N) dx_2 \hdots dx_N \hspace{2pt}.  
\end{equation*}
The OBDM can be expressed in terms of the orbitals of the non-interacting bosons~\cite{BenderErkerGranger2005}: 
\begin{equation}
\label{obdm bender}
\rho_1(x,x\prime)=N\phi_0(x)\phi^*_0(x^\prime)[1-2P(x,x^\prime)]^{N-1}\hspace{2pt},
\end{equation}
where 
\begin{equation}
\label{P bender}  P(x,x^\prime)=\Bigg|\int_x^{x^\prime}|\phi_0(z)|^2dz\Bigg|\hspace{2pt}.
\end{equation}
To find its eigenvalues, commonly referred to as occupation numbers, and associated eigenvectors, known as natural orbitals, one has to solve the following eigenproblem: 
\begin{equation}
\int dx^\prime\rho_1(x,x^\prime)\chi_k(x^\prime)=\lambda_k\chi_k(x)\hspace{2pt}.
\end{equation}
Recently, it has been proven that the eigenproblem can be reduced to a universal form, showing that the occupation numbers are independent of $\phi_0(x)$ and, consequently, completely unaffected by the external potential~\cite{koscik}. 
The demonstration is not limited to the OBDM but is applicable to any $N$-body density matrix.
The universal $\phi$-independent eigenproblem reads~\cite{koscik},
\begin{equation}
\label{phi independent eigenproblem}
\int_0^1 dy^\prime N(1-2|y-y^\prime|)^{N-1}v_k(y^\prime)=\lambda_kv_k(y) \hspace{2pt},
\end{equation}
where the actual eigenvectors $\chi_k(x)$ can be obtained from $v_k(y)$ using the transformation
\begin{equation}
\label{eigenvectors transformation}
\chi_k(x)=\phi_0(x)v_k\big[F(x)\big],
\end{equation}
where $F(x)=\int_{-\infty}^x|\phi_0(z)|^2dz$.
The universal eigenproblem~(\ref{phi independent eigenproblem}) was solved for the simplest system of $N=2$ in Ref.~\cite{koscik} finding that all the eigenvalues are doubly degenerate and equal to $\lambda_{k\pm}=8/[\pi(2k-1)]^2$ with corresponding eigenvectors given by 
\begin{eqnarray}
\label{eigenvectors transformation:v_{k+}}
v_{k+}(y)&=&\sqrt{2}\sin[(2k-1)\pi y],\\
v_{k-}(y)&=&\sqrt{2}\cos[(2k-1)\pi y]. 
\label{eigenvectors transformation:v_{k-}}
\end{eqnarray}
In the following, we extend the solution of the universal eigenproblem for cases where $N>2$. 

By solving the universal eigenproblem~(\ref{phi independent eigenproblem}) for an increasing even number of fermions we find that the eigenvalues always appear in pairs which leads to crucial physical consequences, as will be discussed below. 
We find that the eigenvectors obtained for the $N=2$ case remain valid for larger (and even) numbers of particles.
We obtain the following explicit expressions for eigenvalues (plus-minus sign denotes the double degeneracy)
\begin{equation}
\lambda_{k\pm}= -\frac{384}{[\pi(2k-1)]^4} + \frac{48}{[\pi(2k-1)]^2}
\end{equation}
for $N=4$ and  
\begin{equation}
\lambda_{k\pm}= 
\frac{46080}{[\pi(2k-1)]^6}-\frac{5760}{[\pi(2k-1)]^4}+\frac{120}{[\pi(2k-1)]^2}
\end{equation}
for $N=6$ fermions, while eigenvalues up to $N=10$ are reported in the Supplementary Material. 
By thoroughly examining the analytic expressions for the occupations of the natural orbitals, we have arrived at an explicit analytic expression for the doubly degenerate eigenvalues,
\begin{equation}
\lambda_{k\pm}^{N} = \left\{
\begin{array}{lll}
\sum\limits_{i=1}^{N/2}\frac{(-1)^{i+1}4^i N!}{[(2k-1)\pi]^{2i}(N-2i)!},&\text{even}&N\\
\sum\limits_{i=1}^{(N-1)/2}\frac{(-1)^{i+1}4^i N!}{[2k\pi]^{2i}(N-2i)!},&\text{odd}&N.
\end{array}
\right.
\label{lambda N}
\end{equation}
Expressions~(\ref{lambda N}) constitute the main result of our work and provide valuable insights into the behavior and properties of the system, shedding light on its fundamental characteristics. 
In the absence of an external field, the eigenvalues $\lambda_k$ of the OBDM can be related to the momentum distribution $n(k)$ of a gas on a ring of a circumference $L$, according to $\lambda_k = n(\K)/L$. 
The allowed momenta as $\K = \pm2k\pi/L$ for odd $N$ and $\K =\pm(2k-1)\pi/L$ for even $N$ and Eq.~(\ref{lambda N}) reproduces the momentum distribution found in Ref.~\cite{Sekino2018}.

For an odd number of fermions, all eigenvalues are doubly degenerate except the largest one whose value is always equal to one, $\lambda_0 = 1$, with $v_0(y)=1$ being the corresponding eigenvector.
As for the doubly degenerate eigenvalues, the eigenvectors are slightly different from the ones observed in the even case. 
Specifically, they are given by
\begin{eqnarray}
\label{eigenvectors odd case 1}
v_{k+}(y)&=&\sqrt{2}\sin[2k\pi y],\\
v_{k-}(y)&=&\sqrt{2}\cos[2k\pi y], 
\label{eigenvectors odd case 2}
\end{eqnarray}
and they remain valid for any number of odd particles. 

In particular, the universal eigenproblem~(\ref{phi independent eigenproblem}) can be applied to a plain box with a flat bosonic single-particle state, $\phi_0(x)=1/\sqrt{L}$. 
In that case, function~(\ref{P bender}) expresses as $P(x,x')=|x-x'|/L$ and measures the relative distance.  
In a box of size $L$, OBDM~(\ref{obdm bender}) satisfies $\rho(L,0) = (-1)^{N-1}\rho(0,0)$ which corresponds to periodic (antiperiodic) boundary conditions for odd (even) $N$. 
Such a choice of boundary conditions is appropriate for forming closed shells in a fermionic gas. 
It can be verified, as shown in the Supplementary Material, that the exact eigenstates of the OBDM are given by plane waves $\chi_{k+}(x) = \sqrt{2/L}\cos(\K x)$ and $\chi_{k-}(x) = \sqrt{2/L}\sin(\K x)$ with allowed momenta, i.e. $\K = 0, \pm 2\pi/L, \pm 4\pi/L,\dots \pm2k\pi/L$ for odd $N$ and $\K = \pm\pi/L, \pm 3\pi/L,\dots\pm(2k-1)\pi/L$ for even $N$.
The thermodynamic limit $N\to\infty$ is taken on an untrapped system by increasing the periodicity length $L$ at a fixed density $n=N/L$. 
Quite interestingly, the thermodynamic OBDM can be evaluated explicitly and it exhibits an exponential form, $\rho_1(x,x') = n\exp(-2n|x-x'|)$~\cite{BenderErkerGranger2005}, in contrast to the typical power-law behavior found in compressible systems and predicted by the Luttinger liquid~\cite{Cazalilla_2004}. 
Its Fourier transform provides the momentum distribution $n(\K) = L/[1 + \left(\K/(2n)\right)^2]$ which has a Lorentzian shape. 
The relation between the eigenvalues and the momentum distribution of the untrapped FTG, $\lambda_k = n(\K)/L$, allows for a concise approximation for the eigenvalues as 
\begin{equation}
\lambda_{k\pm}^{N} \approx \left\{
\begin{array}{lll}
\frac{1}{1~+~\left(\pm\pi (k-1/2)\;/\;N\right)^2},&\text{even}&N\\
\frac{1}{1~+~\left(\pm\pi k\;/\;N\right)^2},&\text{odd}&N
\end{array}
\right.
\label{occupation numbers:asymptotic}
\end{equation}
asymptotically decaying as $1/k^2$ for $k\to\infty$ as also discussed in Refs.~\cite{Cui2016,Sekino2018}.

We have conducted numerical verification to validate the correctness of the derived expressions for $\lambda_{k\pm}$ across various values of $N$. 
The numerical results confirm the accuracy and reliability of the analytical findings. 

\begin{figure}[t]
\centering
\includegraphics[width=\columnwidth]{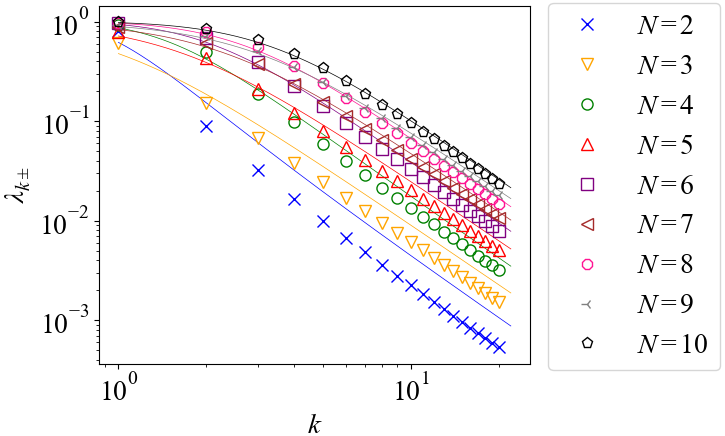}
\caption{Eigenvalues $\lambda_{k\pm}$ as a function of the pair number $k$ for different number of fermions $N$ with symbols. The approximation to Lorentzian shape, see Eq.~(\ref{occupation numbers:asymptotic}), is shown with a continuous line for each number of fermions.}
\label{fig: lambda}
\end{figure}

Figure~\ref{fig: lambda} reports the universal values of the occupations (shown with symbols) of the natural orbitals $\lambda_{k\pm}$ for the different number of fermions as compared to the thermodynamic Lorentzian shape~(\ref{occupation numbers:asymptotic}), shown with lines. 
We find that for $N\gtrsim 10$, the Lorentzian shape is quite precise although it fails in a few-body system.

\begin{figure}[t]
\centering
\includegraphics[width=1\columnwidth]{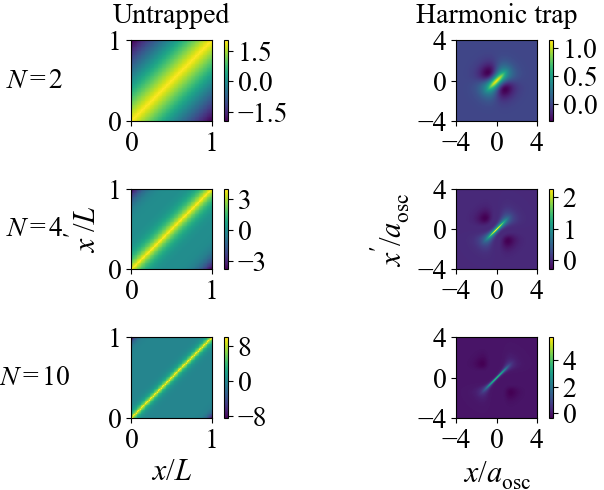}
\caption{OBDM for $N=2$, $N=4$ and $N=10$ for the case of the untrapped case, left column, and a harmonic trap, right column.}

\label{fig: matrices}
\end{figure}

It is crucial to bear in mind that although the occupation numbers $\lambda_k$ are universal and are independent of the specific shape of the external potential~\cite{koscik}, the matrices and corresponding eigenvectors are strongly influenced by the type of external potential used. 
To show that, we report in Fig.~\ref{fig: matrices} characteristic examples of the OBDM $\rho_1(x,x')$ of an untrapped FTG gas and a FTG gas in a harmonic oscillator.
The diagonal terms $n(x) = \rho_1(x,x)$ provide the density profile of the system which is flat in the untrapped case and has a Gaussian shape in a harmonic oscillator. 
The antidiagonal terms $\rho_1(x,-x)$ quantify the loss of coherence with $x\to\infty$ asymptotic value equal to zero (i.e. ODLRO is absent). 

\textbf{Composite boson formation.}  
The formation of composite bosons is evident from the pairing observed in the occupation numbers of the OBDM. In the case of an even number of fermions, all occupation numbers exhibit a double degeneracy whereas for odd $N$ all fermions form composite bosons except one, whose eigenvector coincides with the single-particle ground state with unit eigenvalue. 
To gain a deeper insight into the pairing mechanism, it is instructive to explicitly construct an anti-symmetric two-particle state from pairs of eigenvectors of a doubly degenerate eigenvalue, denoted as $\chi_{k+}(x)$ and $\chi_{k-}(x)$,:
\begin{equation}
\label{slater fermions} 
\psi_{k}(x_1,x_2)=\frac{\chi_{k+}(x_1)\chi_{k-}(x_2)-\chi_{k+}(x_2)\chi_{k-}(x_1)}{\sqrt{2}}\hspace{2pt}.
\end{equation}
The probability of finding a fermion at position $x$ occupying the paired state is identical for both fermions forming the composite boson. 
We denote this probability as $P_{k}(x_1=x)=P_{k}(x_2=x)=P_{k}(x)$ and it is given by 
\begin{equation}
\label{Pij}
P_{k}(x)=\frac{|\chi_{k+}(x)|^2+|\chi_{k-}(x)|^2}{2}\hspace{2pt}.
\end{equation}
This quantity physically corresponds to the density profile of the composite boson with index $k$.
Mathematically, it is formed from two contributions $|\chi_{k+}(x)|^2$ and $|\chi_{k-}(x)|^2$, each dependent according to Eq.~(\ref{eigenvectors transformation}) on the bosonic density profile $|\phi_0(x)|^2$ and two corresponding eigenvectors $v_{k+}(y)$ and $v_{k-}(y)$ defined in Eqs.~(\ref{eigenvectors transformation:v_{k+}}-\ref{eigenvectors transformation:v_{k-}}) and Eqs.~(\ref{eigenvectors odd case 1}-\ref{eigenvectors odd case 2}). 
The crucial aspect is that potential-dependent function $F(x)$ appears as an identical argument in both eigenvectors, $|v_{k+}(F(x))|^2$ and $|v_{k-}(F(x))|^2$, resulting in a summation of the squares of a cosine and sine functions, which sum up to unity.
Consequently, all composite bosons have exactly the same density profile given by
\begin{equation}
\label{p phi}
P_{k}(x)=|\phi_0(x)|^2\hspace{2pt}. 
\end{equation} 
Note that, any linear combination of the orthonormal eigenvectors defined in Eqs.~(\ref{eigenvectors transformation:v_{k+}}-\ref{eigenvectors transformation:v_{k-}}) for the even case and Eqs.~(\ref{eigenvectors odd case 1}-\ref{eigenvectors odd case 2}) for the odd case, would yield exactly the same density of a fermionic pair. Therefore, the density of each of the composite bosons is independent of the freedom of choice of the eigenvectors that one has since the eigenvalues are doubly degenerated. 
\begin{figure}[t]
\centering
\includegraphics[width=1\columnwidth]{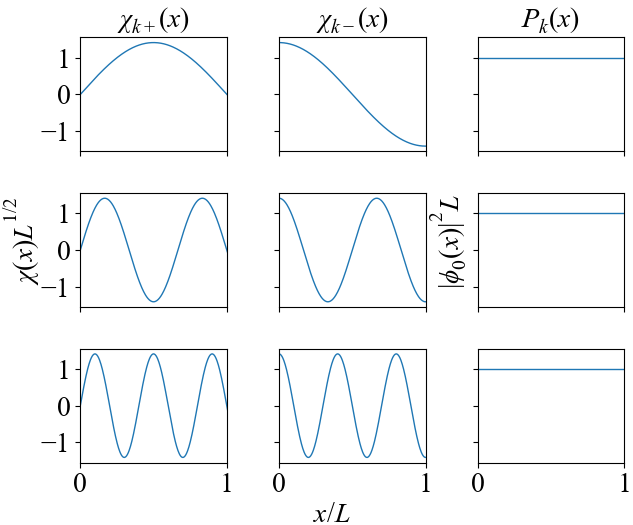}
\caption{
Fermionic natural orbitals and composite boson density profile of an untrapped FTG gas for an even number of particles.
First and second columns, the first six natural orbitals, $\chi_{k+}(x)$ and $\chi_{k-}(x)$, respectively, corresponding to the three largest doubly degenerate eigenvalues $k=1,2,3$ of the OBDM.
Third column, the first three composite boson density profiles $P_{k}(x)$.
}
\label{fig: periodic eigenvectors}
\end{figure}

The total density sums from identical contributions arising from composite bosons, that is $2|\phi_0(x)|^2$ per pair, and a single boson contribution of $|\phi_0(x)|^2$ in the case of an odd number of particles. Consequently, the total density is given by
\begin{equation}
n(x)=N|\phi_0(x)|^2    
\end{equation}
which is equivalent to the density profile of an ideal Bose gas. Then, it follows that given a pair of natural orbitals with the same eigenvalue $k$, if one of them is occupied the other is also occupied forming a composite boson with the contribution to the total density of the system equal to $2|\phi_0(x)|^2$ and the total density of the system is equivalent to that of a Bose-Einstein condensate. Indeed, $n(x)$ being a local quantity, is not affected by the Girardeau mapping and remains the same in FTG and ideal Bose gas.

\begin{figure}[t]
\centering
\includegraphics[width=1\columnwidth]{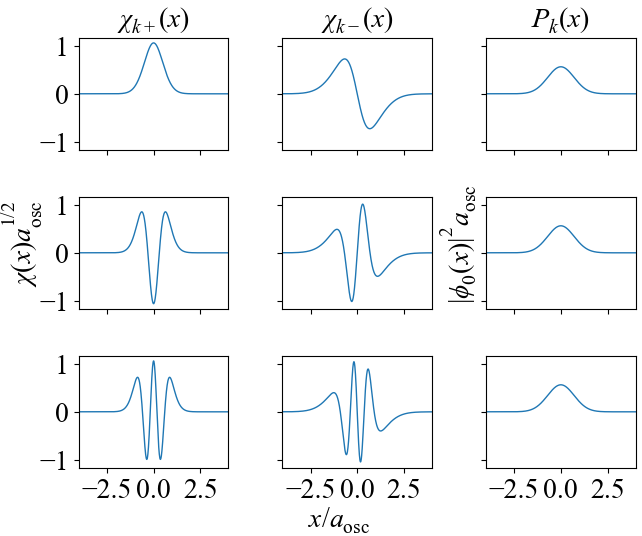}
\caption{
Fermionic natural orbitals and composite boson density profile in a harmonic trap potential for an even number of particles.
First and second columns, the first six natural orbitals, $\chi_{k+}(x)$ and $\chi_{k-}(x)$, respectively, corresponding to the three largest doubly degenerate eigenvalues $k=1,2,3$ of the OBDM.
Third column, the first three composite boson density profiles $P_{k}(x)$.
}
\label{fig: harmonic eigenvectors}
\end{figure}

To gain further insight into the mechanism of composite bosons formation, we thoroughly examine two distinct examples of external potential. 
In the first example, we consider an untrapped FTG gas with flat bosonic density, $\phi_0(x)=1/\sqrt{L}$, in a ring of perimeter $L$. 
The second example involves a harmonic oscillator with the single-particle ground state wave function given by a Gaussian,
\begin{equation}
\phi_0(x)=\frac{1}{\pi^{1/4}\sqrt{a_{\rm osc}}}e^{-x^2/(2a_{\rm osc}^2)},
\end{equation}
where $a_{\rm osc}=\sqrt{\hbar/(m\omega)}$ is the harmonic oscillator length, $m$ the particle mass and $\omega$ the frequency of the trap. 
In Figs.~\ref{fig: periodic eigenvectors}-\ref{fig: harmonic eigenvectors} we show the first six atomic natural orbitals alongside the density of the first three composite bosons in both the untrapped FTG and in the harmonic trap, respectively. 

Notably, while all atomic orbitals differ, as expected for fermions, the density profiles of composite bosons remain consistently the same and equal to $P_{k}=|\phi_0(x)|^2$ for any $k$. 
This provides a real-space picture of how occupied natural orbitals, i.e. fermions, effectively pair to result in a total density equivalent to that of the ideal Bose gas. 

In the context of the formation of composite bosons, the situation recalls the BEC-BCS crossover in two-component Fermi gases. Both systems involve attractive fermion pairing, leading to the formation of composite bosonic pairs.
The key difference lies in molecular structure: In the BCS-BEC crossover, each molecule contains one spin-up and one spin-down fermion with $s$-wave interaction, resulting in the internal structure which is the same for all molecules. On the opposite, in a single-component Fermi gas with $p$-wave interactions, the Pauli exclusion principle applies leading to a much more intricate scenario, as each composite boson has a unique internal configuration while sharing exactly the same density.

\textbf{Conclusions.} 
In this study, we investigate the coherence properties of a fermionic Tonks-Girardeau gas in the presence of an external potential. 
We present an analytical expression~(\ref{lambda N}), for the eigenvalues of the one-body density matrix, applicable for any number of fermions $N$ and under an arbitrary external field. 
For a large number of fermions (in practice, $N\gtrsim 10$) the eigenvalues of the OBDM approach a Lorentzian shape~(\ref{occupation numbers:asymptotic}).
A remarkable feature of the obtained expressions is that they are universal in the sense that the occupation numbers depend only on $N$ and remain independent of the specific shape of the external potential.
The natural orbitals exhibit double degeneracy, implying that pairs of attractive fermions create composite bosons.
In turn, each composite boson exhibits a particle density profile equivalent to that of an ideal Bose gas. This physical picture implies that degenerated natural orbitals must always be either both occupied or both unoccupied. Finally, this pairing results in the fact that the total density of the system is equivalent to that of an ideal Bose gas. 

These findings emphasize the significance of composite bosons formation.
The universality of the results and their independence from the external potential shed light on quantum phenomena occurring in $p$-wave fermions, such as pairing or strong correlations, and pave the way for future explorations in the field of cold atoms, nuclear physics and condensed matter physics.\\

\begin{acknowledgments}
We acknowledge helpful and insightful discussions with Joachim Brand that significantly contributed to the development of this Letter.\\

This work has been funded by Grants No.~PID2020-114626GB-I00 and PID2020-113565GB-C21 by MCIN/AEI/10.13039/5011 00011033 and 
"Unit of Excellence Mar\'ia de Maeztu 2020-2023” 
award to the Institute of Cosmos Sciences, Grant CEX2019-000918-M funded by MCIN/AEI/10.13039/501100011033. 
We acknowledge financial support from the Generalitat de Catalunya (Grants 2021SGR01411 and 2021SGR01095). 
A.R.-F. acknowledges funding from MIU through Grant No. FPU20/06174.
\end{acknowledgments}

\bibliography{biblio}
\bibliographystyle{apsrev4-2}

\appendix
\widetext{
\section{Supplementary Material}
\section{Explicit expressions for the eigenvalues of the OBDM}
\boldmath
\noindent
{\bf Even $N$}\\ 
\unboldmath

\begin{eqnarray}
N=2\;, &\quad&    \lambda_{k\pm}=\frac{8}{[\pi(2k-1)]^2} \\
N=4\;,&\quad& 
    \lambda_{k\pm}=48\Bigg[-\frac{8}{[\pi(2k-1)]^4} + \frac{1}{[\pi(2k-1)]^2}\Bigg] \\
N=6\;,&\quad& \lambda_{k\pm}= 120\Bigg[\frac{384}{[\pi(2k-1)]^6} - \frac{48}{[\pi(2k-1)]^4}+ \frac{1}{[\pi(2k-1)]^2}\Bigg]\\
N=8\;, &\quad& 
\lambda_{k\pm}= 224\Bigg[-\frac{46080}{[\pi(2k-1)]^8}+\frac{5760}{[\pi(2k-1)]^6} - \frac{120}{[\pi(2k-1)]^4}+ \frac{1}{[\pi(2k-1)]^2}\Bigg]\\
N=10\;,&\quad& 
\lambda_{k\pm}= 360\Bigg[\frac{10321920}{[\pi(2k-1)]^{10}}-\frac{1290240}{[\pi(2k-1)]^8}+\frac{26880}{[\pi(2k-1)]^6} - \frac{224}{[\pi(2k-1)]^4}+ \frac{1}{[\pi(2k-1)]^2}\Bigg]
\end{eqnarray}

\boldmath
\noindent
{\bf Odd $N$}\\ 
\unboldmath
\begin{eqnarray}
N=3\;,&\quad&    \lambda_{k\pm}=\frac{24}{[\pi2k]^2} \\
N=5\;,&\quad& 
    \lambda_{k\pm}=80\Bigg[-\frac{24}{[\pi2k]^4} + \frac{1}{[\pi2k]^2}\Bigg] \\
N=7\;,&\quad& 
\lambda_{k\pm}= 168\Bigg[\frac{1920}{[\pi2k]^6} - \frac{80}{[\pi2k]^4}+ \frac{1}{[\pi2k]^2}\Bigg] \\
N=9\;,&\quad& 
\lambda_{k\pm}= 288\Bigg[-\frac{322560}{[\pi2k]^8}+\frac{13440}{[\pi2k]^6} - \frac{168}{[\pi2k]^4}+ \frac{1}{[\pi2k]^2}\Bigg]
\end{eqnarray}

\section{Eigenstates of the OBDM in a box of size $L$}
\boldmath
\noindent
{\bf Even $N$}\\ 
\unboldmath
 The pairs of eigenstates with eigenvalue $\lambda_{k\pm}$ of the OBDM in a box of size $L$ are 
\begin{eqnarray}
\chi_{k+}(x)&=&\sqrt{\frac{2}{L}}\sin\Bigg[\frac{(2k-1)\pi x}{L}\Bigg],\\
\chi_{k-}(x)&=&\sqrt{\frac{2}{L}}\cos\Bigg[\frac{(2k-1)\pi x}{L}\Bigg]. 
\end{eqnarray}
Defining plane waves $\psi_{k+}(x)$ and $\psi_{k-}(x)$ as 

\begin{eqnarray}
\psi_{k+}(x)&\equiv&\frac{1}{\sqrt{2}}\big(\chi_{k+}+i\chi_{k-}\big)=\frac{1}{\sqrt{L}}e^{i(2k-1)\pi x/L},\\
\psi_{k-}(x)&\equiv&\frac{1}{\sqrt{2}}\big(\chi_{k+}-i\chi_{k-}\big)=\frac{1}{\sqrt{L}}e^{-i(2k-1)\pi x/L}, 
\end{eqnarray}
 and since the eigenproblem to be solved is linear and $\chi_{k+}(x)$ and $\chi_{k-}(x)$ share the same eigenvalue $\lambda_{k\pm}$, we prove that the plane waves $\psi_{k+}(x)$ and $\psi_{k-}(x)$ are also eigenstates with eigenvalue $\lambda_{k\pm}$. The plane waves just defined have allowed momenta $\K=\pm(2k-1)\pi/L$ . \\ 

\boldmath
\noindent
{\bf Odd $N$}\\ 
\unboldmath
The prove for an odd number of fermions is really similar but now the eigenstates with eigenvalue $\lambda_{k\pm}$ of the OBDM are
\begin{eqnarray}
\chi_{k+}(x)&=&\sqrt{\frac{2}{L}}\sin\Bigg[\frac{2k\pi x}{L}\Bigg],\\
\chi_{k-}(x)&=&\sqrt{\frac{2}{L}}\cos\Bigg[\frac{2k\pi x}{L}\Bigg]. 
\end{eqnarray}
Thus, the defined plane waves are  
\begin{eqnarray}
\psi_{k+}(x)&\equiv&\frac{1}{\sqrt{2}}\big(\chi_{k+}+i\chi_{k-}\big)=\frac{1}{\sqrt{L}}e^{i2k\pi x/L},\\
\psi_{k-}(x)&\equiv&\frac{1}{\sqrt{2}}\big(\chi_{k+}-i\chi_{k-}\big)=\frac{1}{\sqrt{L}}e^{-i2k\pi x/L}, 
\end{eqnarray}
which, by the same reasoning as in the even case, are also eigenstates of the OBDM with eigenvalue $\lambda_{k\pm}$. The plane waves in the odd case have allowed momenta $\K=\pm2k\pi/L$ .
}
\end{document}